# Application in the Ethanol Fermentation of Immobilized Yeast Cells in Matrix of Alginate/Magnetic Nanoparticles, on Chitosan-Magnetite Microparticles and Cellulose-coated Magnetic Nanoparticles

Viara Ivanova[1], Petia Petrova[1], Jordan Hristov[2]

**Abstract** – *Saccharomyces cerevisiae* cells were entrapped in matrix of alginate and magnetic nanoparticles and covalently immobilized on magnetite-containing chitosan and cellulose-coated magnetic nanoparticles. Cellulose-coated magnetic nanoparticles with covalently immobilized thermostable α-amylase and chitosan particles with immobilized glucoamylase were also prepared. The immobilized cells and enzymes were applied in column reactors – 1/for simultaneous corn starch saccharification with the immobilized glucoamylase and production of ethanol with the entrapped or covalently immobilized yeast cells, 2/ for separate ethanol fermentation of the starch hydrolysates with the fixed yeasts. Hydrolysis of corn starch with the immobilized α-amylase and glucoamylase, and separate hydrolysis with the immobilized α-amylase were also examined. In the first reactor the ethanol yield reached approx. 91% of the theoretical; the yield was approx. 86% in the second. The ethanol fermentation was affected by the type of immobilization, the initial particle loading, feed sugar concentration and the dilution rate. The ethanol productivity with entrapped cells reached 264.0 g/L.h at particle loading rate 70% and dilution rate 3.0 $h^{-1}$ with reducing sugar concentration of 200.0 g/L. The prepared magnetic particles with fixed yeast cells were stable at 4°C in saline for more than 1 month.

## I. Introduction

Bioethanol is the most widely used liquid biofuel. Its market is expected to reach 100 $\times 10^9$ liters in 2015 [1]. The largest producers in the world are the United States, Brazil, and China. In 2009, the US produced $39.5 \times 10^9$ liters of ethanol using corn as a feedstock while the second largest producer, Brazil, created about $30 \times 10^9$ liters of ethanol using sugarcane. China is a country that has invested much in the production of ethanol, and is nowadays one of the largest ethanol producers [2]. Currently, almost all bio-ethanol is produced from grain or sugarcane ("first generation" of ethanol production) [3]-[4].

Fermentation systems operated in continuous mode offer a number of advantages compared to batch processes, generally resulting in enhanced volumetric productivity and, consequently, smaller bioreactor volumes and lower investment and operational costs [5]. These continuous processes can benefit from whole cell immobilization techniques in order to retain high cell densities inside the bioreactors. The immobilized biocatalysts have been extensively investigated during last few decades. Immobilization of microbial cells showed certain technical and economical advantages over free cell system [6].

The immobilization methods can be classified into 4 categories [6]-[7]: carrier-binding, cross-linking, entrapping, and a combination of these 3 methods. Lack of toxicity and economic feasibility should be considered in food technologies. Several natural polysaccharides, such as alginates, κ-carrageenan, agar, and agarose, are excellent gel materials and are widely used for entrapment [8]. Many other support materials for cell immobilization have been reported including delignified cellulosic material, chitosan [9]-[10], natural zeolite [11], g-alumina etc.

Perspective techniques for yeasts immobilization can be divided into four categories: attachment or adsorption to solid surfaces (wood chips, delignified brewer's spent grains, DEAE cellulose, and porous glass), entrapment within a porous matrix (calcium alginate, k-carrageenan, polyvinyl alcohol, agar, gelatine, chitosan, and polyacrilamide), mechanical retention behind a barrier (microporous membrane filters, and microcapsules) and self-aggregation of the cells by flocculation. The application of these different immobilization methodologies and carriers, their impact in microbial growth and physiology, internal and external mass transfer limitations, product quality and consistency, bioreactor design, bioprocess engineering and economics have been largely discussed [12]-[13]-[14]-[15]-[16]. Several theories have been proposed to explain the enhanced fermentation capacity of microorganisms as a result of immobilization. A reduction in the ethanol concentration in the immediate microenvironment of the organism due to the formation of a protective layer or specific adsorption of ethanol by the support may act to minimize end product inhibition [17]-[18].

In brewing, industrial continuous systems using immobilized cells have been successfully implemented for the production of alcohol-free beer and for beer maturation [19]-[20]-[21]. Entrapment of microbial cells like yeast helped the brewing industry to reduce fermentation process times and the size of their storage facilities [22]. The most significant advantages of immobilized yeast cell systems are the ability to operate with high productivity at dilution rates exceeding the maximum specific growth rate, the increase of ethanol yield and cellular stability and the decrease of process expenses due to the cell recovery and reutilization [17]. Other advantages of immobilized cell system over presently accepted batch or continuous fermentations with free-cells are: greater volumetric productivity as a result of higher cell density; tolerance to higher concentrations of substrate and products; lacking of inhibition; relative easiness of downstream processing etc. in different types of bioreactors, such as packed bed reactor, fluidized bed reactor, gas-lift reactor and reactor with magnetic field [18]-[23]-[24].

Continuous fermentation processes based on immobilized microbial cells have the potential to maximize the volumetric productivity while minimizing the production costs in the bio-ethanol industry. Nevertheless, the industrial experience with continuous systems for the ethanol fermentation has been mostly limited to pilot scale demonstration plants [19]-[20].

Despite of the slow incorporation by the industry, research in the field of continuous bio-ethanol fermentation with immobilized cells has been quite active [5]. The development and testing of new carrier materials for microbial immobilization and application in continuous fermentations is also an active research topic. Novel supports recently described include natural materials such as brewer's spent grains [14] or sorghum bagasse [25], as well as synthetic materials like microporous divinyl benzene copolymer [26] or magnetic particles for use in a magnetically stabilized fluidized bed reactor [27]-[28].

Yeasts, particularly *Saccharomyces* spp., are usually the choice for industrial ethanol production, because of their good fermentative capacity, high tolerance to ethanol and other inhibitors (either formed during raw-materials pre-treatments or produced during fermentation) and the capacity to grow rapidly under the anaerobic conditions that are characteristically established in large-scale fermentation vessels. The exploitation of microorganisms other than *Saccharomyces cerevisiae* for continuous alcoholic fermentations has also been reported, namely *Zymomonas mobilis* entrapped into polyvinyl alcohol hydrogel [29], engineered *Escherichia coli* immobilized on porous glass beads [30]-[31] and the thermophilic anaerobic bacterium *Thermoanaerobacter* strain BG1L1 immobilized on granular carrier material for the fermentation of glucose and xylose in undetoxified lignocellulosic hydrolysates [32]-[33].

The aim of this study was to immobilize hydrolytic enzymes and yeast cells in matrix of alginate/magnetic nanoparticles, on chitosan-magnetite microparticles and cellulose-coated magnetic nanoparticles and to apply these immobilized biopreparations for corn starch hydrolysis and ethanol fermentation of the obtained hydrolysates.

## II. Experimental

### II.1. Yeast strain

The strain *Saccharomyces cervisiae* C12 (NBIMCC, Bulgaria) was maintained on YPG agar medium (pH 4.5), containing yeast extract (3.0 g/L), peptone (5.0 g/L), glucose (10.0 g/L), agar (20.0 g/L). Further propagation, prior to immobilization and ethanol fermentation, was performed in 500 ml flasks in a bath or in incubator at 30°C for 48 h in a defined liquid media. The liquid media consisted of yeast extract (3.0 g/L), peptone (3.5 g/L), $KH_2PO_4$ (2.0 g/L), $MgSO_4 \times 7H_2O$ (1.0 g/L), $(NH_4)_2SO_4$ (1.0 g/L), glucose (10 g/L). The yeast suspension of a desired cell density was prepared from the yeast slurry after centrifugation of the cultured media.

### II.2. Enzymes

Termamyl SC, a heat-stable α-amylase from *Bacillus licheniformis* (Novozymes, Denmark) was used for corn starch liquefaction. The enzyme activity was 133 KNU/g (KNU, kilo novo units of α-amylase is the amount of enzyme which breaks down 5.26 g of starch per hour according to Novozymes standard method for the determination of α-amylase).

Spirizyme Plus FG (*Aspergillus niger* glucoamylase, Novozymes, Denmark) with specific activity 400 AGU/g (AGU is the amount of enzyme which hydrolyses 1 μmol of maltose per minute under specified conditions) was used for corn starch saccharification.

### II.3. Immobilization

#### II.3.1. Immobilization of yeasts in alginate magnetic beads

The immobilization in calcium alginate magnetic beads (CAMB) was carried out as follows: The yeast cell slurry and the sodium alginate solutions were mixed with dried magnetic nanoparticles (prepared by standard precipitation of Fe(II) and Fe(III) with molar ratio of 1:2 in alkaline medium, [34]-[35]-[36]) to 3% (w/v) final concentration of sodium alginate, final cell content 100-250 mg/mL gel and 5-20% (w/v) final nanoparticles content. The mixture obtained was extruded dropwise into gently stirred and chilled 2% (w/v) $CaCl_2 \cdot H_2O$ and kept for 2 h for stabilization. The calcium alginate magnetic beads containing the cells were thoroughly washed with distilled water and used for further studies.

### II.3.2. Immobilization of yeasts on chitosan-magnetite microparticules

The yeast cells were immobilized on chitosan-magnetite microparticles (CHMM) by covalent bonding using glutaraldehyde as cross-linking and activating agent [9]. Powdered chitosan (Fluka, Mw $7.5 \cdot 10^5$) was dissolved in 2% acetic acid. Into 20 mL of 1% chitosan solution, containing 50-150 mg of magnetite powder (Iron (II, III) oxide, $Fe_3O_4$, Aldrich, diameter < 5 μm), 13.2 mL of 0.5M KOH was added gradually at 50°C under stirring. After 10 minutes 0.5 g of glutaraldehyde (from 25% solution, (Fluka)) was added. Stirring was continued for additional 30 minutes. Chitosan/magnetite microparticles were washed with 0.1 M Na-acetate buffer, pH 5.0. A solution of yeast cells (250 mg/g particles) was added and the mixture was stirred at 20°C for 30 min and then left at 4°C overnight. Next day the particles were washed until no cells were detected in the washes. Immobilized yeasts were stored at 4°C in 0.1 M Na-acetate buffer, pH 5.0.

### II.3.3. Immobilization on cellulose-coated magnetic nanoparticles

The cellulose-coated magnetic nanoparticles (CCMN) were obtained by coagulation of cellulose solution containing magnetic nanoparticles in ammonium sulfate [37]. 8 g of cellulose, 12 g of NaOH and 8 g of urea were added to distilled water. The suspension was put at 4°C for 12-18 h and then stirred vigorously at 10°C to obtain almost transparent solution. To this aqueous solution of cellulose, pre-calculated quantity of magnetite microparticles (25-35% final concentration, w/v) was added under constant stirring to ensure homogeneous mixing. The suspension was dropped into 8% (w/v) ammonium sulfate solution under vigorous stirring. The resulting microparticles were filtered, washed several times with distilled water and then dried at 60°C.

A 5 g quantity of cellulose-coated magnetite microparticles was put in 200 mL of 0.05M periodic acid. The pH of the solution was adjusted to 4.0, followed by heating in water bath at 80°C under constant stirring for 15 h. The particles were then filtered, washed with distilled water and then dried at 40°C.

A 1.0 g quantity of dry oxidized CCMN particles was re-suspended in 0.1 M phosphate buffer, pH 6.0, containing 250 mg of yeast biomass and incubated under stirring for 90 min and then at 4°C overnight. The particles with immobilized cells were washed with distilled water and 0.1 M Na-acetate buffer solution, pH 5.0.

### II.3.4. Immobilization of enzymes

To 10 mL buffered α-amylase solution were added 2.5 g of cellulose-coated magnetic nanoparticles. The mixture was stirred at 4°C, 25°C or 40°C for 2-18h. The enzyme quantity in the solution was from 0.5 to 2.5 mg/mL. After the coupling step, the supports were washed with an abundant quantity of water until no protein was detectable in the filtrates, determined by the method of Bradford [38] at 595 nm. Bovine serum albumin (BSA) was used as a standard protein.

Immobilization of glucoamylase on activated with glutaraldehyde chitosan-magnetite microparticles was carried out similarly at 4°C, 15°C or 22°C for 2-18h. The samples were washed with distilled water to remove the unbound enzyme and then stored at 4°C.

### II.4. Reactor configuration

Two reactors with different sizes and geometries were built to carry continuous processes. Jacketed reactor (R1) was used for starch saccharification with immobilized glucoamylase and had a total volume of 0.5 L and working volume of 0.45 L. The inner tube diameter was 4.0 cm, and the height/diameter ratio was 10. The temperature was maintained at 60°C with water from a water bath. Peristaltic pumps were used to pump in liquefied starch medium and to re-circulate hydrolysed fluid within the reactor.

The second reactor (R2) had a total volume of 1.0 L, working volume of 0.95 L, inner tube diameter was 5.0 cm, and the height/diameter ratio was approx. 10. The reactor was inclined up to 45°. Thus the weight of the immobilized cell particles was distributed on larger surface and the released $CO_2$ traversed a smaller column of beads and medium. The fermentations in this reactor were carried out at 30°C. Peristaltic pumps were used to pump in fermentation medium and to withdraw the fluid from the fermentor. The reactors and tubes were sterilized with sodium hypochlorite solution (10%, v/v) and rinsed with sterilized water prior to operation. The gas outlets at the top of the reactors were protected with sterile cotton wool filters, to reduce the risk of contamination.

*II.5. Corn starch liquefaction, saccharification and fermentation*

Batch separate liquefaction of 30.0% non-gelatinized corn starch with the immobilized α-amylase was performed in 500 mL flasks in water bath with agitation (Julabo SW22, Germany, 150 rpm) at 90°C and pH 6.5. Biocatalyst particles reached from 20% to 50% of reaction volume (150 mL). 0.05 mg of $CaCl_2$ per g starch was added prior to liquefaction for stabilization and activation of the enzyme. Immobilized amylase was separated from the liquefied starch using a magnet.

Batch saccharification was performed at 60°C with agitation and pH 4.0 (adjusted after the liquefaction). The volume of immobilized enzyme particles was from 20% to 50% of the total volume (150 mL). Continuous saccharification with recirculation was carried out in the jacketed column reactor (R1) at 60°C. The volume of the chitosan-magnetite microparticles with immobilized glucoamylase (Vb) was from 40 to 60% of the reactor volume (Vf), and the dilution rate in the continuous process was from 0.05 to 0.4 $h^{-1}$.

Simultaneous saccharification of 6% corn starch with the immobilized glucoamylase and production of ethanol with the entrapped or covalently immobilized yeast cells was carried out at 30°C. The inclined reactor was filled up to 80% of its volume with immobilized glucoamylase and immobilized yeast cells (1:1). The starch was gelatinized with boiling before pumping.

Separate ethanol fermentation of the starch hydrolysates with the fixed yeasts was carried out continuously in the inclined reactor (R2) at 30°C. The particles with immobilized *Saccharomyces cerevisiae* cells reached from 30 to 70% of reactor volume. The bioreactor was run batch-wise for 8 h, to establish the yeast biomass prior to starting continuous culture operation.

In both fermentation experiments pH was not controlled.

The starch hydrolysates were autoclaved at 121°C for 15 minutes. The fermentation medium was supplemented before sterilization with $(NH_4)_2SO_4$ (2.0 g/L).

*II.6. Analytical methods*

*II.6.1. Determination of the amount of bound enzyme*

The amount of protein loaded onto the supports was determined using a modified Bradford assay. This method measures the decrease in the absorbance of the solution at 465 nm due to adsorption of the dye by the bound protein. For this assay, 200 μL of distilled water was added to 4.5 mg of enzyme loaded support and 4.5 mg of blank support, respectively. Then 1800 μL of Bradford reagent, diluted (1:5) with distilled water, was subsequently added to each of these mixtures. After short agitation of the mixture (3 min) to allow binding of the dye with the protein, the mixtures were centrifuged at 3000 g. The absorbance of the two supernatants was measured at 465 nm. The reading for the enzyme loaded support was then subtracted from the similar reading obtained with the blank support. The calibration curve for quantification of the protein was obtained with BSA as a reference protein at 465 nm. The binding (BY) yield was determined as follows: BY, [%] = (P/Po) x100, where Po and P are the protein added and the measured after immobilization, as [mg protein]. The specific enzyme activity was defined as: Specific enzyme activity = Enzyme activity/mg of protein. The activity yield (AY) was calculated as a ratio of the specific enzyme activity measured after immobilization [U/mg protein] to the specific enzyme activity before the immobilization [U/mg protein] and was expressed as percentage.

*II.6.2. Enzyme assay*

The starch-degrading activity was measured by the method of Pantchev *et al*. [39]. The assay is based on the decrease in absorbance of the iodine-amylose complex as a result of soluble starch degradation. One unit of amylase activity was defined as the amount of enzyme causing hydrolysis of 0.162 mg of soluble starch to dextrins per min under defined assay conditions (1.0% Lintner starch, 0.066M phosphate buffer, pH 6.5, 30°C).

Glucoamylase activity was assayed by the quantity of liberated glucose from soluble starch. The glucose was determined using the Miller's method [40]. One unit of glucoamylase activity was defined as the amount of enzyme which liberates 1 μmol of glucose per minute from 1% soluble starch at pH 4.0 and 60°C.

*II.6.3. Immobilized enzyme assay*

About 10.0 mg (wet) carriers with immobilized enzyme and 4.0 ml of 1.0% soluble starch (pH 6.5, 0.066M phosphate buffer (amylase) or pH 4.0, 0.5M phosphate-citrate buffer (glucoamylase) were incubated at 30°C or 60°C for 15 min. 1.0 mL of reaction mixture was taken for enzyme assay.

*II.6.4. Other assays*

The samples were centrifuged immediately after collection, and then the supernatant was filtered through 0.2 μm pore size membrane filter (Acrodisc syringe filters, Pall, UK) and frozen for later analysis.

Glucose concentrations of diluted samples were assayed using Miller's method [40]. Ethanol of diluted samples was determined by Ethanol Assay Kit A-111(Biomedical Research Service, SUNY Buffalo, NY, US). The test is based on the reduction of the tetrazolium salt INT in a NADH-coupled enzymatic reaction to formazan, which is water-soluble and exhibits an absorption maximum at 492 nm. The intensity of the red color formed increased in the presence of increased ethanol concentrations (up to 0.5%).

Dry weight of the biomass: Yeast cells were harvested by centrifugation for 10 min at 5000 rpm. They were washed twice with distilled water and dry weight was measured by Ultra X2010 automatic analyzer, Germany.

## III. Results and Discussion

*III.1. Enzyme binding and coupling efficiency*

Covalently immobilized α-amylase Termamyl SC on cellulose-coated magnetic nanoparticles and glucoamylase Spirizyme Plus FG, immobilized on chitosan via glutaraldehyde were prepared. The results for the immobilization efficiency as a function of the enzyme quantity in the solution, the temperature and the reaction time are summarized in Tables I-III.

TABLE I
PROTEIN BINDING AND COUPLING EFFICIENCY OF COVALENTLY IMMOBILIZED AMYLASE AND GLUCOAMYLASE

| Enzyme | Protein Added (mg/mL support) | | | Binding yield (%) | | | Activity yield (%) | | |
|---|---|---|---|---|---|---|---|---|---|
| **1.** Termamyl SC | 2.0 | 5.0 | 10.0 | 83 | 70 | 63 | 67 | 63 | 55 |
| **2.** Spirizyme Plus FG | 2.0 | 5.0 | 10.0 | 92 | 76 | 58 | 78 | 75 | 52 |

Preparations with 1.8-6.3 mg/mL carrier bound α-amylase were obtained. Similar were the results for the glucoamylase – the immobilized protein reached 1.92-5.1 mg/mL carrier. Higher activity - up to 4850 U/mL support were obtained at amylase concentration in the immobilization solution 2.0 mg/mL. The higher binding yield (BY) 83-92% was reached at the lowest protein concentration – 2.0 mg/mL for both enzymes. At enzyme concentrations 5.0 mg and 10.0 mg/mL the corresponding binding yields were lower.

Higher binding yields and satisfactory activity yields were obtained when the immobilization process was carried out at 4°C (glucoamylase) and at 25°C (amylase).

The maximum binding yields were obtained after 18 hours of coupling (Table III). In this experiment the immobilization solutions contained 5 mg protein per mL support. The maximum glucoamylase activity yields were obtained after 2 h of reaction at low binding yields. The results indicate that the immobilization rate of this enzyme on the support is very fast at low enzyme concentration in the immobilization solution. Similar effects on the amylase activity were not proved.

The activity yield (AY) reflects an activity loss due to the immobilization of enzyme and was calculated on the base of enzyme specific activity before and after the fixation. The inactive part of the fixed enzymes was probably due to changes in the enzyme structure or steric, microenvironmental and diffusional effects, as it was observed for immobilized amylases [41]-[42]-[43].

TABLE II
EFFECT OF THE TEMPERATURE ON THE PROTEIN LOADING AND THE ACTIVITY YIELD; TERMAMYL SC (1) AND SPIRIZYME PLUS FG (2)

| Temperature (ºC) | Bound enzyme (mg protein/mL support) | Binding yield (%) | Activity yield (%) |
|---|---|---|---|
| 4 | (1) 1.35 | (1) 67.5 | (1) 58 |
| 4 | (2) 1.84 | (2) 92 | (2) 78 |
| 25 | (1) 1.66 | (1) 83 | (1) 67 |
| 15 | (2) 1.7 | (2) 85 | (2) 73 |
| 40 | (1) 1.47 | (1) 73.5 | (1) 65 |
| 22 | (2) 1.55 | (2) 77.5 | (2) 71 |

TABLE III

EFFECT OF THE REACTION TIME ON THE PROTEIN LOADING AND THE ACTIVITY YIELD OF PREPARATIONS; TERMAMYL SC (1), AND SPIRIZYME PLUS FG (2)

| Time (h) | Bound enzyme (mg protein/mL support) | Binding yield (%) | Activity yield (%) |
|---|---|---|---|
| 2 | (1) 1.45 | (1) 29 | (1) 67 |
|   | (2) 1.65 | (2) 33 | (2) 86 |
| 6 | (1) 2.85 | (1) 57 | (1) 70 |
|   | (2) 2.7 | (2) 54 | (2) 78 |
| 18 | (1) 3.5 | (1) 70 | (1) 63 |
|   | (2) 3.8 | (2) 76 | (2) 75 |

*III.2. Corn starch liquefaction and saccharification*

The α-amylases are endohydrolases which cleave 1,4-α-D-glucosidic bonds and can bypass but cannot hydrolyse the 1,6-α-D-glucosidic branchpoints. Commercial enzymes used for the industrial hydrolysis of starch are produced by *Bacillus amyloliquefaciens* (supplied by various manufacturers) and by *Bacillus licheniformis* (supplied by Novozymes as Termamyl).

In this study the non-gelatinized corn starch liquefaction was done at 90°C. The hydrolysis in batch process in flasks and the effect of biocatalyst volume are shown on Fig. 1. As a result of the high enzyme concentration used per gram substrate, the corn starch was hydrolyzed to dextrose equivalents (DE) 18-20 after 60 minutes of action.

The dextrose equivalent represents the percentage hydrolysis of the glycosidic linkages. Starch has a DE of zero. During starch hydrolysis, DE indicates the extent to which the starch has been cleaved. The maximum DE obtainable using bacterial α-amylases is around 40. In most commercial processes where the starch is subjected to further saccharification, the DE values of 8-12 are usually used. The principal requirement for liquefaction is to reduce the viscosity of the gelatinised starch to facilitate subsequent processing.

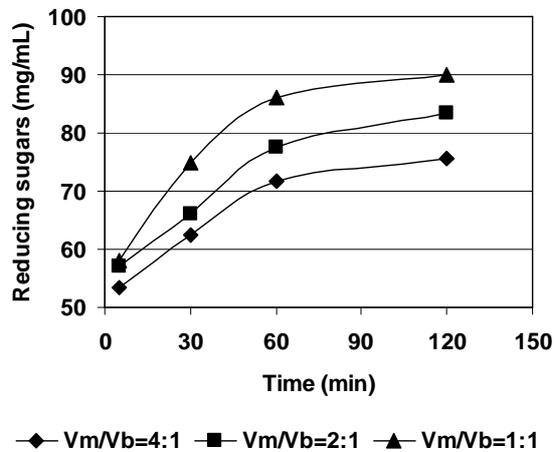

Fig.1. Hydrolysis of corn starch with immobilized amylase Termamyl SC and effect of biocatalyst volume in batch process.

After batch saccharification with immobilized glucoamylase the reducing sugars reached more than 200 mg/mL in 18 hours (Fig. 2). In commercial processes the dosage of glucoamylase is usually 0.2 U/g starch and the DE could reach 95-97 in 72 hours of batch process. In our experiments the immobilized glucoamylase in the flasks was from 0.47 U/g starch (Vm:Vb=4:1) to 1.9 U/g starch (Vm:Vb=1:1). Dextrose equivalents (DE) of 88-90 were obtained after 42 hours of saccharification at 60°C. The biocatalyst volume in the flasks was not proved to be a significant factor for the process effectiveness because of the higher enzyme quantity per gram of starch used.

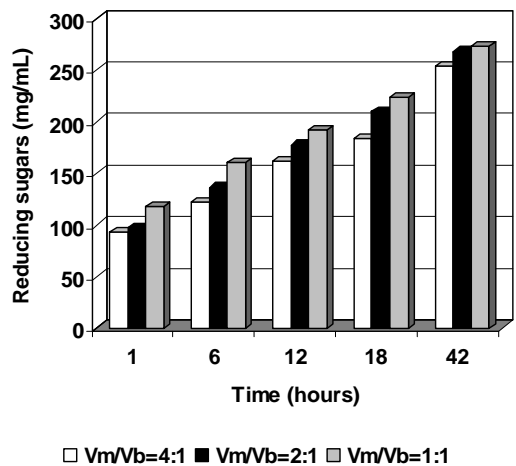

Fig.2. Batch saccharification with immobilized glucoamylase Spirizyme Plus FG - effect of biocatalyst volume.

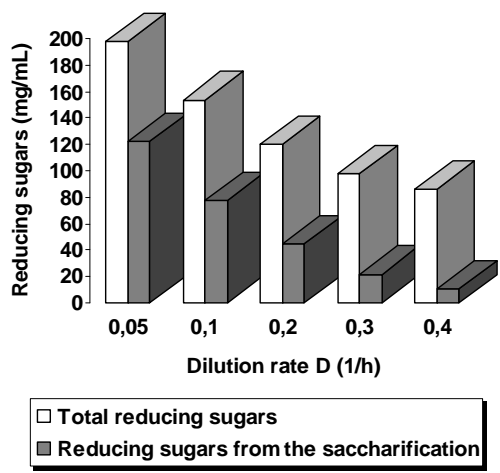

Fig.3. Continuous saccharification with immobilized glucoamylase – effect of the dilution rate (Vm:Vb=2:3).

Continuous saccharification in the column reactor and the effect of the dilution rate are presented on Fig. 3. Starch hydrolysates, obtained with immobilized amylase Termamyl SC at medium/biocatalyst ratio Vm/Vb=4:1 were used. The initial reducing sugar concentration was 75.7 mg/mL. The volume of chitosan-magnetite microparticles with immobilized glucoamylase was from 40 to 60% of the reactor volume, and the dilution rate in the continuous process was from 0.05 to 0.4 $h^{-1}$. The glucoamylase concentration was from 1.26 U/g starch to 2.85 U/g starch, i.e. from 6 to 14 times higher than in conventional industrial process. The results showed that the process could be carried out continuously at low dilution rates. Saccharification process was effective at D 0.05 $h^{-1}$ and the DE reached was 65, i.e. 65% of the starch was hydrolyzed to maltodextrins after the liquefaction and saccharification steps. At higher dilution rates (D 0.4 $h^{-1}$) the impact of the saccharification in the total reducing sugar content decreased sharply to only 10 mg/mL.

From the results for continuous saccharification with recirculation (Fig. 4) it is evident that better results could be obtained in this process configuration. The degree of hydrolysis reached 96% after 40 hours of operation compared to 90% in batch experiments and to 65% in continuous process at dilution rate 0.05$h^{-1}$. After 40 hours at maximal desired DE the process could be stopped and the biocatalyst could be used in next process.

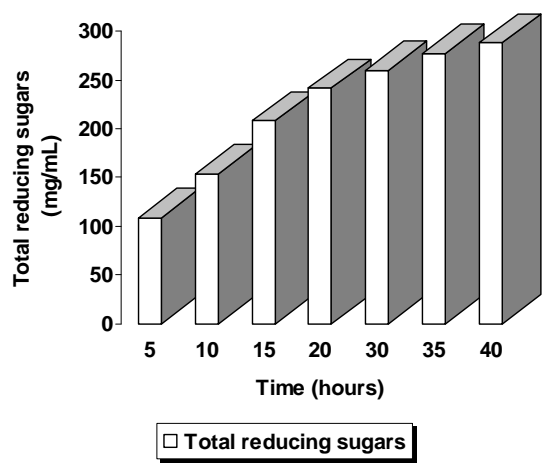

Fig.4. Continuous saccharification with recirculation (Vm:Vb=2:3; D=0.05$h^{-1}$; Vfresh medium:Vrecirculated=1:1).

In industry liquefaction is usually a continuous process but saccharification is most often conducted as a batch process. The saccharification process takes about 72 h to complete but may, of course, be speeded up by the use of more enzyme. Continuous saccharification is possible and practicable if at least six tanks are used in series. The reaction is stopped by heating to 85°C for 5 min, when a maximum DE has been attained. Further incubation results in a fall in the DE, to about 90 DE, eventually caused by the formation of isomaltose as accumulated glucose re-polymerises with the approach of thermodynamic equilibrium [44]-[45].

*III.2. Ethanol fermentation of liquefied and saccharified starch*

*III.2.Ethanol fermentation with immobilized yeast cells*

Yeast cells were immobilized in calcium alginate gel, containing magnetic nanoparticles (calcium alginate magnetic beads (CAMB)), on chitosan-magnetite microparticles (CHMM) by covalent bonding using glutaraldehyde as cross-linking and activating agent, and covalently on oxidized to dialdehyde cellulose-coated magnetic nanoparticles (CCMN).

Alginate pellets with initial concentration of the immobilized biomass 100-250 mg/mL were studied. Nevertheless, significant differences between the results for ethanol yield on substrate glucose were not observed. During the fermentation process part of the substrate is used for cell development and maintaining of cell viability. The biomass concentration in the pellets reaches very fast higher, and approx. equal concentrations, so the next experiments were done with the beads containing initially 100 mg/mL gel immobilized yeast cells.

*III.2.2. Separate ethanol fermentation of the starch hydrolysates with the fixed yeasts*

Ethanol productions were investigated in inclined column reactor with working volume 950 mL at 30°C. Figures 5, 6 and Table IV summarize the results obtained with the three types of biocatalysts. The starch hydrolysates were diluted with water to reducing sugar concentrations 120.0 g/L and 200.0 g/L, supplemented with $(NH_4)_2SO_4$ and sterilized. Volume of alginate pellets was from 285 to 665 mL (30, 55, 65 and 75% of the working volume). Volumes of chitosan-magnetite microparticles and cellulose-coated magnetic nanoparticles were 285 and 380 ml (30 and 40% of the working volume). The particle loading rate was calculated as a ratio of loading particle volume to the bioreactor working volume.

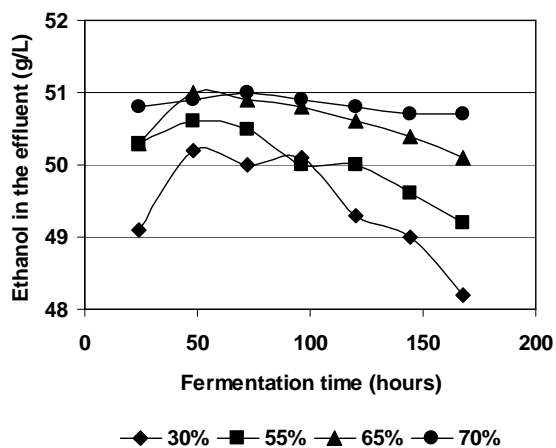

Fig.5. Continuous fermentation of starch hydrolysates with immobilized in calcium alginate magnetic beads (CAMB) yeast cells- effect of the particle loading (reducing sugars 120 g/L).

The ethanol theoretical yield, productivity and concentration reached 83%, 71.3 g/L h, and 51.0 g/L, respectively, at a particle loading rate of 65% (v/v) and a feed dilution rate of 1.4 $h^{-1}$ with an initial reducing sugar concentration of 120.0 g/L. At 70% particle loading rate (v/v) and initial reducing sugars 200.0 g/L the respective values were: 86% for the theoretical yield, 88.0 g/L for the ethanol, and 264.0 g/L.h for the productivity at dilution rate 3.0 $h^{-1}$. The beads with immobilized in alginate magnetic particles *Saccharomyces cervisiae* C12 cells were relatively stable without significant reduction in activity for 168 hours of continuous fermentation.

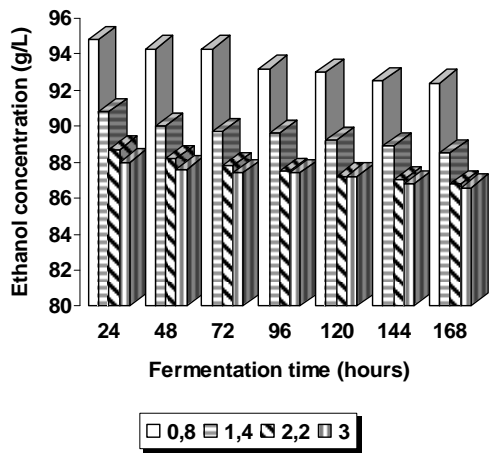

Fig.6. Continuous fermentation of starch hydrolysates with immobilized in calcium alginate magnetic beads (CAMB) yeast cells - effect of the dilution rate (particle loading rate 70%; initial reducing sugars 200.0 g/L).

Free-cell concentrations were tested. They increased from 0.08 (12 h) to 1.01 g/L at the end of fermentation (168 h). This increase could be due to the growth of free cells leaking from gel beads.

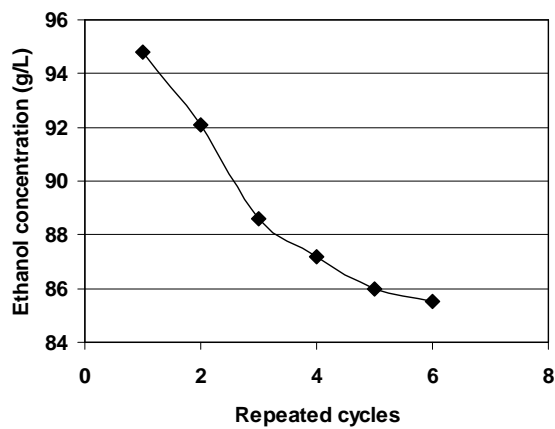

Fig.7. Repeated use of immobilized in alginate cells in continuous processes.

To investigate the fermentation efficiency of immobilized cells after storage the particles were separated by magnet, washed with distilled water and buffer and stored at 4°C for two months in saline. Every 10 days they were applied in a new continuous process for 168 h, then washed and stored again. The results are presented on Fig. 7. The prepared magnetic particles with fixed yeast cells were stables; the cells retained their viability and fermented the hydrolysates to higher ethanol concentrations – from approx. 94 g/L in the first continuous process to 85 g/L in the sixth. In total, the reactor with immobilized *S. cerevisiae* cells was run successfully for more than 42 days without a significant loss of ethanol productivity.

TABLE IV
ETHANOL FERMENTATION OF STARCH HYDROLYSATES
(1) – REDUCING SUGARS 120 G/L; (2) - REDUCING SUGARS 200 G/L; PARTICLES LOADING 70%; DILUTION RATE 1.4 H$^{-1}$.

| Immobilization matrices | Ethanol (g/L) | Ethanol yield (g/g sugar) | Percentage of the theoretical yield (%) |
|---|---|---|---|

| Immobilization matrices | Ethanol (g/L) | Ethanol yield (g/g sugar) | Percentage of the theoretical yield (%) |
|---|---|---|---|
| **1.** CAMB | (1) 50.8<br>(2) 90.8 | (1) 0.4233<br>(2) 0.4540 | (1) 82.8<br>(2) 88.8 |
| **2.** CHMM | (1) 45.7<br>(2) 73.6 | (1) 0.3808<br>(2) 0.3680 | (1) 74.5<br>(2) 72.0 |
| **3.** CCMN | (1) 41.3<br>(2) 74.8 | (1) 0.3442<br>(2) 0.3740 | (1) 67.3<br>(2) 73.1 |

The performance of ethanol fermentation of starch hydrolysates with covalently immobilized on chitosan-magnetite microparticles and on cellulose-coated magnetic nanoparticles was also affected by initial particle loading rate, feed sugar concentration and dilution rate and the results are shown on Table IV. The ethanol concentration and productivity increased with increasing particle loading rate from 30% (v/v) to 40% (v/v). The ethanol yield as g/g sugar and the percentage of the theoretical yield decreased as the sugar concentration and the feed dilution rate increased in the continuous fermentation process with the immobilized on chitosan-magnetite microparticles (CHMM) *S. cerevisiae* cells. The ethanol productivity in the inclined reactor with immobilized *S. cerevisiae* cells increased with the increase of sugar concentration and dilution rate, but the ethanol concentration decreased with the increase of dilution rate. This phenomenon was also observed in experiments with immobilized *S. cerevisiae* cells using other carriers [12]-[18]-[46]-[47].

*III.2.1. Simultaneous corn starch saccharification with the immobilized glucoamylase and production of ethanol with the entrapped or covalently immobilized yeast cells*

Ethanol yield based on consumed sugars and percentages of the theoretical yield are summarized in Table V. The continuous ethanol fermentation was investigated at a particle loading rate of 30-40% (v/v) and at dilution rates of 0.1 h$^{-1}$ and 0.3 h$^{-1}$.

TABLE V
SIMULTANEOUS ETHANOL FERMENTATION AND STARCH SACCHARIFICATION
(1) – DILUTION RATE 0.1 H$^{-1}$; (2) - DILUTION RATE 0.3 H$^{-1}$; PARTICLE LOADING 80% (FIXED GLUCOAMYLASE/FIXED YEASTS=1:1); REDUCING SUGAR CONCENTRATION 36.4 G/L (1), 30.2 G/L (2), 33.8 (3), 30.8 (4), 32.6 (5), 29.5 (6).

| Immobilization matrices | Ethanol (g/L) | Ethanol yield (g/g sugar) | Percentage of the theoretical yield (%) |
|---|---|---|---|
| **1.** CAMB | (1) 16.9<br>(2) 13.9 | (1) 0.4653<br>(2) 0.4603 | (1) 91.0<br>(2) 90.0 |
| **2.** CHMM | (3) 15.6<br>(4) 14.0 | (3) 0.4603<br>(4) 0.4545 | (3) 90.0<br>(4) 88.9 |

| Immobilization matrices | Ethanol (g/L) | Ethanol yield (g/g sugar) | Percentage of the theoretical yield (%) |
|---|---|---|---|
| 3. CCMN | (5) 14.9<br>(6) 13.3 | (5) 0.4570<br>(6) 0.4508 | (5) 89.4<br>(6) 88.2 |

Immobilized glucoamylase hydrolysed the 6% corn starch at 30°C to DE 40-60. The process was slow and was carried out 72 hours. The maximal reducing sugar concentration obtained was 36.4 g/L at dilution rate of 0.1 $h^{-1}$. The sugars were fermented to ethanol with yield approx. 90-91% of the theoretical, but the ethanol concentration and the productivities were low, due to the low substrate concentration. Free cells were observed and their concentration at the end of fermentation was from 0.9 (CHMM) to 1.7 g/L (CCMN). This increase could be due to the growth of free cells detached from the supports. When the cells are covalently immobilized the daughter cells could not be retained on the particles.

On Table VI some data from the literature are compared with the obtained results in this study. It could be concluded that the application of inclined column reactor, high dilution rates and particles loading rates, and immobilized yeast cells achieves ethanol productivity comparable or higher than that in other investigations.

TABLE VI
COMPARATIVE DATA FOR ETHANOL FERMENTATION WITH IMMOBILIZED *SACCHAROMYCES CEREVISIAE* IN VARIOUS BIOREACTOR SYSTEMS.

| Reactor | Substrate | Dilution rate (1/h) | Productivity (g/L.h) | Yield (%) | Ref. |
|---|---|---|---|---|---|
| A | Molasses | 0.2 | 24.4 | 62.0 | [48] |
| B | Molasses | 0.4 | 16.0 | 70.0 | [46] |
| C | Glucose | 0.37 | 21.8 | 96.0 | [18] |
| D | Starch hydroly-sates | 1.4<br>3.0 | 90.8<br>264.0 | 88.8<br>86.0 | This work |

A - Column packed bed reactor; B - fluidized bed reactor; C - magnetically stabilized fluidized bed reactor; D – inclined column reactor.

## IV. Conclusion

Industrial starch conversion includes three stages: - 1/ gelatinization, involving the dissolution of the nanogram-sized starch granules to form a viscous suspension; - 2/ liquefaction, involving the partial hydrolysis of the starch, with concomitant loss in viscosity; and 3/ saccharification, involving the production of glucose and maltose by further hydrolysis. Gelatinization is achieved by heating starch with water. Gelatinised starch is readily liquefied by partial hydrolysis with enzymes and saccharified by further enzyme hydrolysis.

Enzymes used in starch hydrolysis are bacterial α–amylases (*Bacillus amyloliquefaciens, B. licheniformis*) and saccharifying α–amylase (*B. subtilis*), α–amylase from *Aspergillus oryzae* and *Aspergillus niger*, β-amylase, glucoamylase from *Aspergillus niger* and bacterial pullulanase.

The amylase from *B. licheniformis* can cleave only α-1,4-oligosaccharide links to give α-dextrins and predominantly maltose, G3, G4 and G5 oligosaccharides. With the glucoamylase from *A. niger* the α-1,4 and α-1,6-links are cleaved from the nonreducing ends, to give β-glucose.

Using *Saccharomyces* yeasts in traditional batch fermentations for distilled ethanol production; productivity is limited to only 1.8–2.3 g /L.h, which is uneconomic for biofuel production. Although continuous fermentation can increase this rate, higher rates can be achieved if cell retention is also employed. In conventional industrial processes separation of cells from the product stream can be achieved using gravity - [49] or centrifuge-assisted [50] sedimentation or membrane separation [51] and recycle, or immobilization within the fermenter. This is the Brazilian solution, where continuous centrifuges are used to recycle biomass from the spent broth. However, centrifugation is expensive, in terms of both capital and running costs. Separation and recycle requires additional equipment and energy consumption and is therefore less suitable for the manufacture of marginal-cost products such as renewable fuels. In contrast, immobilization of cells does not require cell separation and recycle.

In this study, an effective method for immobilization of the yeast *Saccharomyces cervisiae* C12 was applied and the entrapped cells produces high levels of ethanol for more than 42 days. A suitable immobilization and process configuration was applied for liquefaction and saccharification of corn starch with the obtained immobilized biocatalysts – a thermostable commercial α–amylase and glucoamylase.

An inclined bed reactor has been applied for continuous ethanol fermentation with the immobilized yeast cells in/on magnetic particles, and the immobilized cells of *S. cerevisiae* possessed the capacity to yield high ethanol productivities during the course of continuous fermentation. The results obtained showed that these immobilized biocatalysts could be used for non-magnetical or magnetical applications in the starch and alcohol industries.

The developed process of continuous saccharification of starch hydrolysates with recirculation offers a possibility for application in industrial processes.

# Authors' information

1. Department of Organic Chemistry and Microbiology, University of Food Technologies, 26, Maritsa Blvd., 4002 Plovdiv, Bulgaria
2. Department of Chemical Engineering, University of Chemical Technology and Metallurgy, Sofia

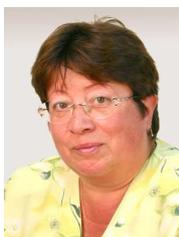

**Viara Ivanova**
vn.ivanova@abv.bg
Associate professor Viara Ivanova received her Ph.D. in Fermentation Chemistry and Biotechnology from the Prague University of Chemical Technology, Czech Republic, in 1986. Her current research is focused on the application of magnetic micro- and nanoparticles for protein immobilization.

Prof. Ivanova is a lecturer in Microbiology in the University of Food Technologies, Plovdiv, Bulgaria.

**Petia Petrova** is a PhD student in the Department of Organic Chemistry and Microbiology of the University of Food Technologies, Plovdiv, Bulgaria.

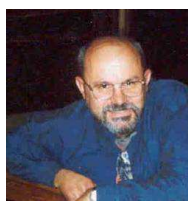

**Jordan Hristov**
jordan.hristov@mail.bg
Jordan Hristov is associate professor of Chemical Engineering at the University of Chemical Technology and Metallurgy, Sofia, Bulgaria. He was graduated in 1979 as Electrical Engineer (MS equivalent) at the Technical University, Sofia, Bulgaria. His

PhD thesis on the magnetically assisted fluidization was awarded by the University of Chemical Technology and Metallurgy in 1995. Prof. Hristov's research interests cover the areas of particulate solids mechanics, fluidisation, heat and mass transfer with special emphasis on scaling and approximate solution. Relevant information is available at http://hristov.com/jordan.